\documentclass[a4paper,12pt]{spieman}

\usepackage{orcidlink}
\usepackage{graphicx}
\usepackage{xcolor}
\usepackage{booktabs}
\usepackage{subcaption}
\usepackage{siunitx}
\usepackage{amsmath}
\usepackage{ulem}
\usepackage{url}
\usepackage{lineno}

\title{X-ray Response of the Fully-Depleted, p-Channel SiSeRO-CCD}

\author[1]{Julian~Cuevas-Zepeda\orcidlink{0000-0002-2358-7049}}
\author[2]{Joseph~Noonan\orcidlink{0000-0002-9702-2518}}
\author[3]{Claudio~Chavez\orcidlink{0000-0002-7853-6900}}
\author[4]{Miguel~Sofo-Haro\orcidlink{0000-0001-9397-2922}}
\author[3]{Nathan~Saffold\orcidlink{0000-0001-6358-9228}}
\author[1,3,5]{Juan~Estrada}
\author[6]{Kevan~Donlon}
\author[6]{Chris~Leitz}
\author[7]{Steve~Holland}

\affil[1]{Department of Astronomy and Astrophysics and the Kavli Institute for Cosmological Physics, University of Chicago, Chicago, IL 60637, USA}
\affil[2]{Department of Physics, University of Chicago, Chicago, IL 60637, USA}
\affil[3]{Fermi National Accelerator Laboratory, Batavia, IL 60510, USA}
\affil[4]{Comisión Nacional de Energía Atómica (CNEA) \& Consejo Nacional de Investigaciones Científicas y Técnicas (CONICET), Universidad Nacional de C\'ordoba, C\'ordoba C5000, Argentina}
\affil[5]{Brookhaven National Laboratory, Upton, NY 11973, USA}
\affil[6]{MIT Lincoln Laboratory, Lexington, MA 02421, USA}
\affil[7]{Lawrence Berkeley National Laboratory, Berkeley, CA 94720, USA}

\begin{document}
\maketitle

\begin{abstract}
We present an X-ray characterization of a fully depleted, 725~$\mu$m thick p-channel SiSeRO-CCD. 
Measurements with a $^{55}$Fe source yield an energy resolution of $54 \pm 0.9$~eV ($14.6 \pm 0.25~e^{-}$) at 5.9~keV for single-pixel events, indicating that the SiSeRO amplifier preserves the intrinsic charge resolution of the CCD under multi-sample non-destructive readout.
Characterization with a $^{241}$Am source extends the response to higher-energy photons, with reconstructed spectral features observed between 9--26~keV and the 59.5~keV $\gamma$ emission. 
These measurements, together with a muon-derived diffusion calibration, show that charge transport and diffusion are consistent with interactions spanning the full sensor depth. 
These results demonstrate that the SiSeRO-CCD simultaneously achieves sub-electron noise performance and efficient charge collection in a thick, fully depleted silicon detector. 
This combination enables X-ray spectroscopy across a broad energy range while maintaining sensitivity to faint signals.
\end{abstract}

\keywords{SiSeRO-CCD, X-ray detectors, CCDs, non-destructive readout, fully depleted silicon, astronomical instrumentation}

\noindent{\footnotesize\textbf{*}Corresponding author: \linkable{juliancz@uchicago.edu}}

\begin{spacing}{2}

\section{Introduction}

The detection and precise measurement of faint optical and X-ray signals are central to many applications in astronomy and low-background instrumentation. 
Charge-coupled devices (CCDs) have long served as the standard imaging detectors for these applications due to their excellent linearity, stability, and low-noise performance.
However, conventional CCD architectures face a fundamental trade-off between readout noise and readout speed, which limits their use in applications that require both high sensitivity and frame rates.

Recent developments in CCD readout architectures have shown that non-destructive charge measurements can significantly reduce effective readout noise by enabling multiple measurements of the same charge packet. 
For example, Skipper CCDs achieve sub-electron read noise by repeatedly sampling the sense node without destroying the stored charge~\cite{tiffenberg2017single}.
While highly effective for ultra-low-noise applications, this approach typically requires long readout times, limiting its use in systems that require higher frame rates.

For astronomical instrumentation, this trade-off is especially important in low-signal regimes where detector read noise, rather than photon shot noise, sets the sensitivity. 
This occurs in ground-based observations at blue wavelengths, where the sky background is low, and in space-based observations, where exposures are often split into many short integrations to mitigate cosmic-ray contamination. 
In both cases, read noise dominates the noise budget and directly limits performance.
Recent work has demonstrated the suitability of Skipper CCDs for astronomical observations, including quantum efficiency and detector response measurements~\cite{drlica2020characterization} and on-sky spectroscopy with a prototype Skipper focal plane~\cite{Villalpando_2024,Marrufo_Villalpando_2024}. 
However, the long readout times required to achieve sub-electron noise with Skipper CCDs limit their practical use in astronomical observations. The time spent on readout reduces the time available for photon collection, limiting observing efficiency, cadence, and scalability to large-format focal planes. 
This motivates the development of alternative readout architectures that preserve low-noise performance while enabling faster charge measurement.

The Single-electron Sensitive ReadOut (SiSeRO) architecture was developed to address this limitation by integrating a double-gate MOSFET amplifier that enables non-destructive readout while improving charge measurement speeds. 
Previous work has demonstrated that the SiSeRO amplifier achieves sub-electron charge sensitivity and can operate several times faster than floating-gate Skipper amplifiers~\cite{2023_prl_sisero, 2023_sisero, 10.1117/12.3020855, 10.1117/1.JATIS.10.1.016004, 10.1117/1.JATIS.9.2.026001, chattopadhyay2022first}.
In a fully depleted thick CCD, this combination is attractive for astronomy because it couples a high-sensitivity output stage with a detector geometry that provides high quantum efficiency and bulk charge collection~\cite{holland2023fully}.

In this work, we evaluate the X-ray response of a fully depleted, p-channel SiSeRO-CCD. 
Using $^{55}$Fe and $^{241}$Am radiation sources, we measure the detector's energy resolution and demonstrate efficient charge collection across the thick substrate. 
These measurements provide an initial characterization of the SiSeRO CCD as a candidate detector for future space-based astrophysics missions, including concepts for the Habitable Worlds Observatory, where low-noise photon detection and efficient bulk charge collection are critical for detecting faint biosignatures.

\section{Methods}
\label{sec:methods}

\subsection{The SiSeRO-CCD}
\label{sec:methods/sisero}

The Single-electron Sensitive Readout (SiSeRO) CCD used in this study is a 725~$\mu$m thick, p-channel CCD fabricated by MIT Lincoln Laboratory.
The sensor consists of 1278 $\times$ 330 pixels with a pixel size of 15 $\times$ 15~$\mu$m$^2$.
It was fabricated on high-resistivity ($\approx$20~k$\Omega\cdot$cm) n-type silicon, which enables full depletion of the device volume with a 70\,V substrate bias~\cite{2023_prl_sisero,2023_sisero}.
The p-channel CCD architecture provides improved resistance to radiation-induced charge transfer degradation~\cite{Cervantes-Vergara_2025,roach2024effectsprotonirradiationperformance,roach2025sdw, dawson2008radiation, 8048486}, while the fully depleted thick substrate extends the sensor’s quantum efficiency into the near-infrared~\cite{holland2003fully,holland2002overview}.
Similar to the Skipper CCD, the SiSeRO-CCD enables non-destructive readout (NDR) and achieves sub-electron noise performance~\cite{tiffenberg2017single, 2023_prl_sisero, 10.1117/12.3020855}.

\begin{figure}[t]
\centering

\begin{subfigure}[t]{0.48\linewidth}
    \centering
    \includegraphics[width=\linewidth]{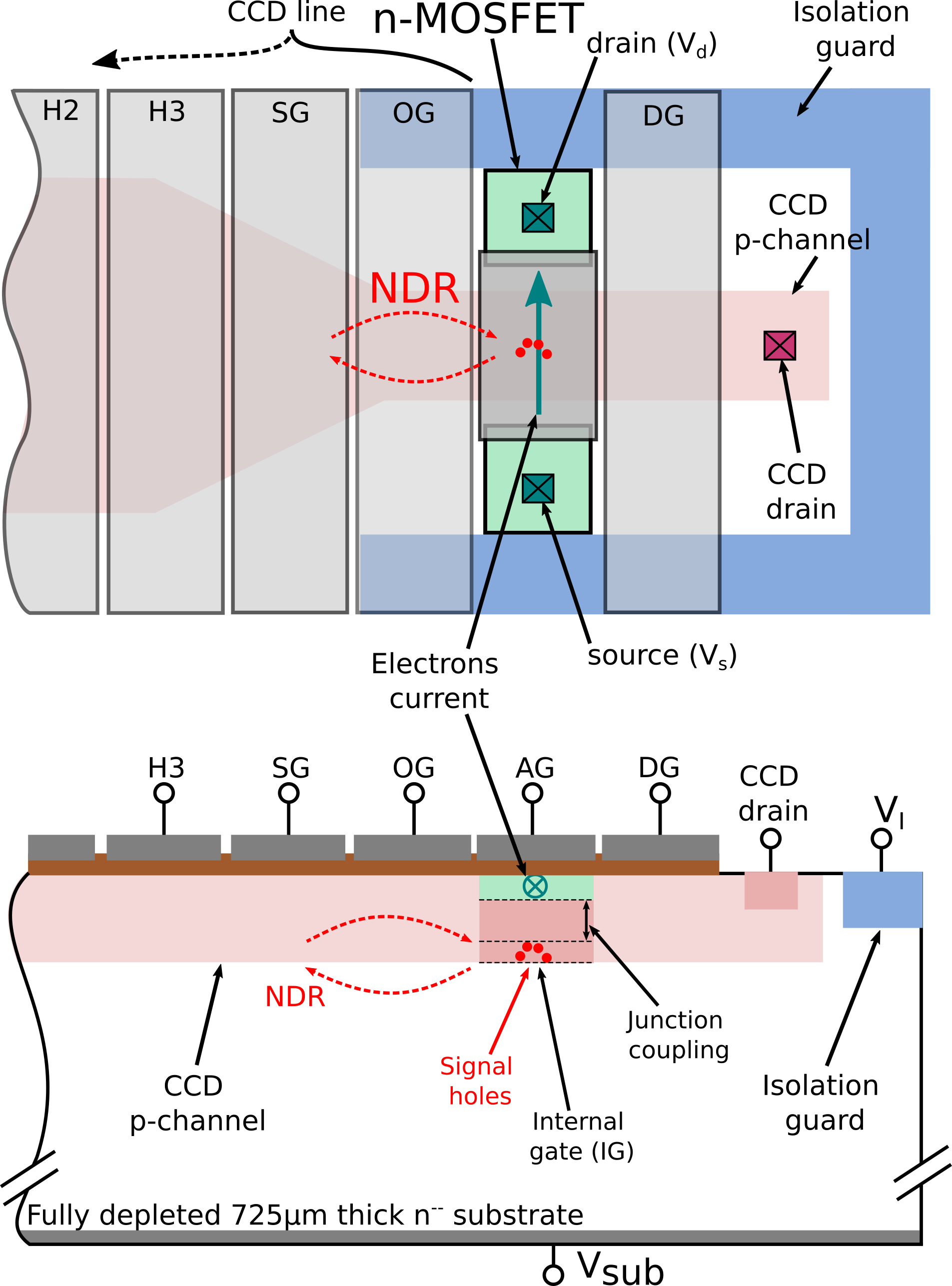}
    \caption{}
    \label{fig:sisero}
\end{subfigure}
\hfill
\begin{subfigure}[t]{0.48\linewidth}
    \centering
    \includegraphics[width=\linewidth]{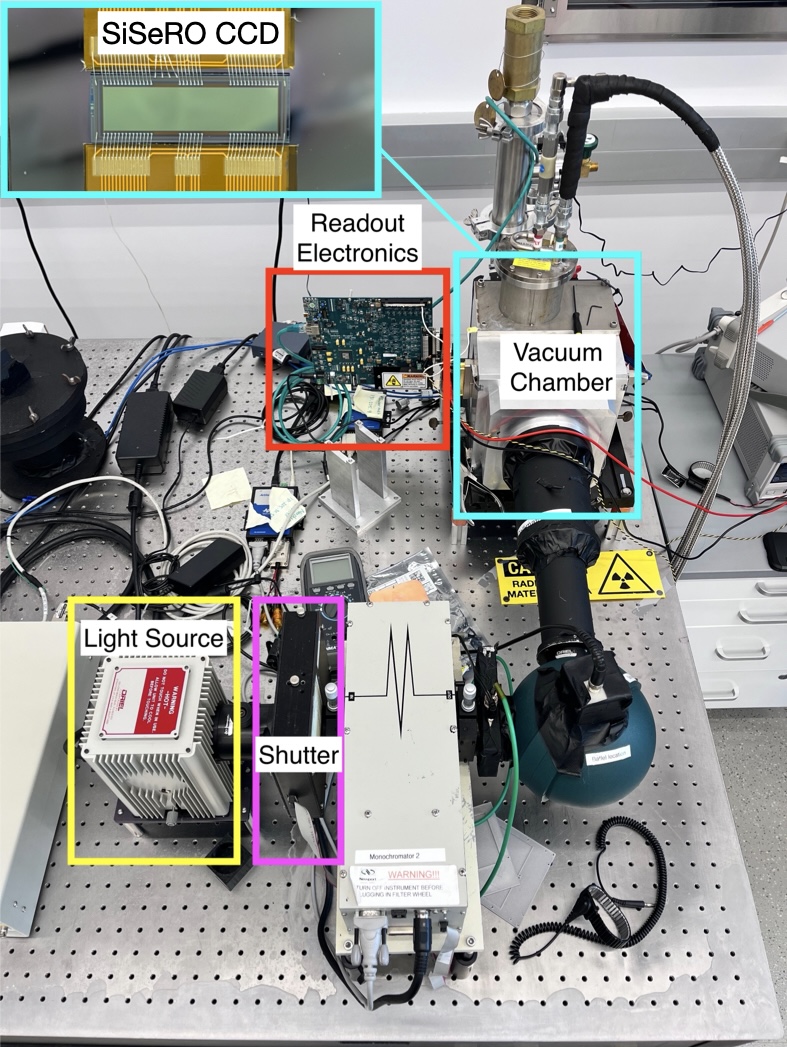}
    \caption{}
    \label{fig:setup}
\end{subfigure}

\caption{
    Overview of the SiSeRO-CCD architecture and experimental setup used for detector characterization. 
    (a) Schematic top and cross-sectional views of the SiSeRO output stage and amplifier, illustrating the integration of an n-MOSFET within the CCD p-channel and the principle of non-destructive readout (NDR). The figure highlights the key CCD clocking elements and amplifier structures, including the summing-well, output, amplifier, and dump gates, as well as the isolation guard and junction-coupled internal gate. 
    (b) Photograph of the experimental setup, including the sensor, vacuum chamber, readout electronics, and optical excitation components.
    }
\label{fig:sisero_setup}

\end{figure}

In this device, charge readout is performed using a double-gate n-type MOSFET amplifier architecture referred to as SiSeRO.
The SiSeRO architecture is depicted in Fig.~\ref{fig:sisero}.
By design, the n-channel MOSFET is integrated into the CCD’s p-channel.
The p-channel functions as an internal gate (IG), forming the second gate of a double-gate structure that couples the CCD charge to the transistor channel.
This geometry establishes a coupling between the stored charge and the drain current $I_{DS}$, allowing the charge packet to modulate the current without being transferred or lost.
As a result, charge packets can be measured non-destructively, and the design yields enhanced charge sensitivity compared to conventional floating-diffusion and floating-gate amplifiers~\cite{2023_prl_sisero, 10.1117/1.JATIS.9.2.026001, 10.1117/12.3020855}.
Device simulations further indicate that operating the MOSFET near its threshold voltage maximizes charge sensitivity, a prediction validated by measurements in 2023 that achieved 1.54 nA/$e^-$, the highest reported for a double-gate MOSFET~\cite{2023_sisero,2023_prl_sisero}.
Additionally, experimental measurements have demonstrated that the SiSeRO architecture achieves equivalent noise performance at least six times faster than the Skipper CCD, confirming its advantage in the speed-to-noise trade-off regime~\cite{2023_prl_sisero, castoldi_experimental_2024}.

To isolate the n-type regions of the SiSeRO amplifier from the n-type substrate of the CCD array, an isolation guard consisting of a p-type implant surrounds the output stage (Fig.~\ref{fig:sisero}). This guard prevents parasitic flow of electrons into the transistor channel, a requirement for stable operation and full device depletion~\cite{2023_prl_sisero}.
Additionally, as has been experimentally observed, the absence of the isolation guard causes a rapid increase in drain-source current during non-destructive readout and prevents the reduction of pixel rms to sub-electron levels.
Simulations and experimental measurements confirm that $-8$~V is required to bias the isolation guard and prevent parasitic current from affecting pixel readout~\cite{2023_sisero,2023_prl_sisero}.

During readout, the amplifier is operated by applying the bias voltages $V_{GS}$ and $V_{DS}$ to drive the output signal $I_{DS}$.
Once the charge packet reaches the serial register, it is clocked to the amplifier for readout.
When charge is transferred into the internal gate, the output gate (OG) is raised to prevent charge from drifting back into the serial register, and the charge modulates $I_{DS}$.
The drain-source current response exhibits a nonlinear dependence on the total charge stored in the internal gate and is consistent with simulations and previous studies of bulk charge detectors that employ a similar output amplifier architecture~\cite{2023_sisero,hynecek1999bcd}.
The front-end electronics employ a transimpedance amplifier (TIA) to convert current variations to a voltage for digitization.
An offset potential $V_{\mathrm{offset}}$ is applied in the TIA to introduce a controlled bias current, ensuring that the resulting output voltage remains within the dynamic range of the readout system.
Under multi-sample readout, the OG is lowered to send charge back into the summing-well gate (SG).
The OG is then raised again, and the pedestal signal is read out.
Subsequently, the charge is returned to the amplifier, read out, and the process is repeated until the desired number of pixel samples is acquired.
Once complete, the dump gate (DG) is lowered, and the pixel charge is drained to allow the next pixel to be read out.

The performance of the SiSeRO output stage strongly depends on the bias configuration and readout electronics~\cite{cuevaszepeda2025automatingsensorcharacterizationbayesian}.
The following section details the experimental setup and operating parameters chosen to evaluate the device's response to X-rays.

\subsection{Experimental Setup}

A series of characterization measurements were conducted on a SiSeRO device at Fermi National Accelerator Laboratory.
The SiSeRO-CCD was installed in an aluminum vacuum chamber and pumped down to $10^{-5}$~Torr.
Using a CryoTiger PT-30 cryocooler, the sensor was cooled to 130~K to reduce dark current.
The Low Threshold Acquisition (LTA) was used as the readout system~\cite{2021_lta}.
An optical ensemble with a light source, a monochromator, an automated mechanical shutter, and an integrating sphere was attached to a viewport in front of the vacuum chamber to uniformly illuminate the sensor during dedicated amplifier optimization runs (Fig.~\ref{fig:setup}).
All radiation sources used for measurements in this paper were installed inside the vacuum chamber, facing the exposed front-illuminated sensor.

\label{sec:methods/params}

The sensor used in this study was previously characterized for sub-electron performance~\cite{2023_prl_sisero}.
The bias parameters used to operate the sensor are summarized in Table~\ref{tab:operating_configs}.
These operating configurations were obtained using a previously developed Bayesian-optimization--based characterization framework~\cite{cuevaszepeda2025automatingsensorcharacterizationbayesian}.

The SiSeRO CCD was operated with a correlated double sampling (CDS) time of $\sim$27~$\mu$s. 
In this configuration, the pixel readout speed was limited by the LTA readout system~\cite{2023_prl_sisero}. 
Images were acquired using either single-sample readout or non-destructive readout (NDR). 
For each exposure, the NDR samples were averaged to form a single image, followed by a row-wise overscan subtraction to remove the electronic baseline.

\begin{table}[t]
\centering
\caption{Operating bias conditions for readout of the SiSeRO-CCD.}
\label{tab:operating_configs}
\begin{tabular}{l S[table-format=1.2] c}
\toprule
\textbf{Parameter} & \textbf{Value} & \textbf{Unit} \\
\midrule
$V_{GS}$               & 4.30  & V \\
$V_{DS}$               & 0.42  & V \\
$V_{\mathrm{offset}}$  & 8.00  & V \\
$V_{\mathrm{ISO}}$     & -8.00 & V \\
\bottomrule
\end{tabular}
\end{table}

\section{X-ray Response and Charge Transport}
\label{sec:energy}

This section presents measurements of the X-ray response and charge collection properties of the SiSeRO-CCD using laboratory X-ray sources and cosmic-ray muons. 
These measurements probe key detector characteristics: charge transport through a thick fully depleted substrate, pixel's charge collection, and charge measurement resolution.

\subsection{Fe-55 Measurements}

To evaluate the energy resolution of the SiSeRO-CCD, a series of exposures were acquired using an $^{55}$Fe source under the operating conditions described in Sec.~\ref{sec:methods/params}, using NDR with 100 samples per pixel.
This measurement focuses on the Mn $K_{\alpha}$ (5.9~keV) and $K_{\beta}$ (6.5~keV) emission lines produced following the decay of $^{55}$Fe.

\begin{figure}[ht]
  \centering
  \includegraphics[width=0.75\linewidth]{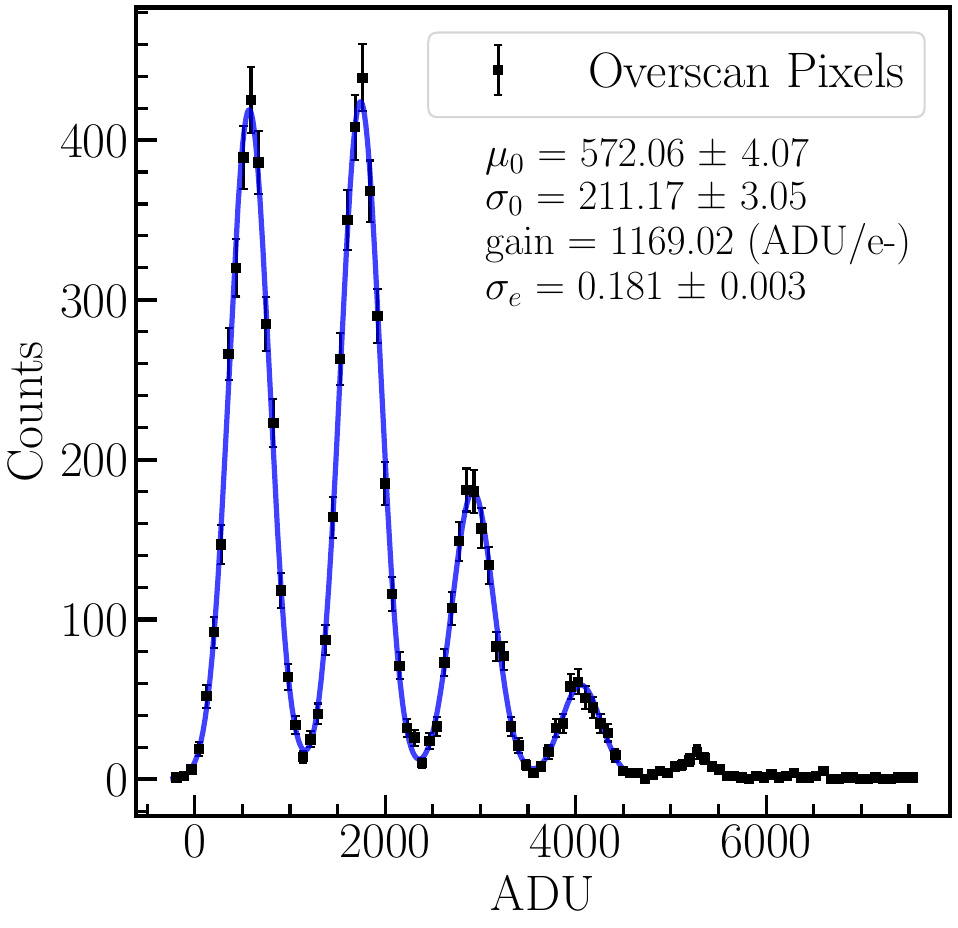}
  \caption{
        Histogram of pixel values in the overscan region, showing discrete charge peaks corresponding to integer numbers of electrons (starting at zero). 
        Black points show the data, and the blue curve shows the best-fit multi-Gaussian model used to extract the gain and read noise. 
        The clear separation of the single-electron peaks demonstrates sub-electron read noise and enables precise conversion between ADU and collected charge.
    }
  \label{fig:overscan-peaks}
\end{figure}

To assess the noise performance of the SiSeRO CCD, we analyzed the pixel value distribution in the overscan region.
The overscan does not contain any illuminated pixels, so all recorded signal arises from the serial register~\cite{haro_measurement_2016}.
Figure~\ref{fig:overscan-peaks} shows the pixel value distribution from the overscan region of a 100-sample NDR image.
The presence of well-resolved multi-electron peaks demonstrates that the SiSeRO amplifier can resolve individual charge carriers.
Fitting the first four peaks (0$e^{-}$, 1$e^{-}$, 2$e^{-}$, 3$e^{-}$) yields a readout noise of $\sigma_{e}=0.181\pm0.003~e^{-}$.

\begin{figure}[h]
    \centering
    \includegraphics[width=0.75\linewidth]{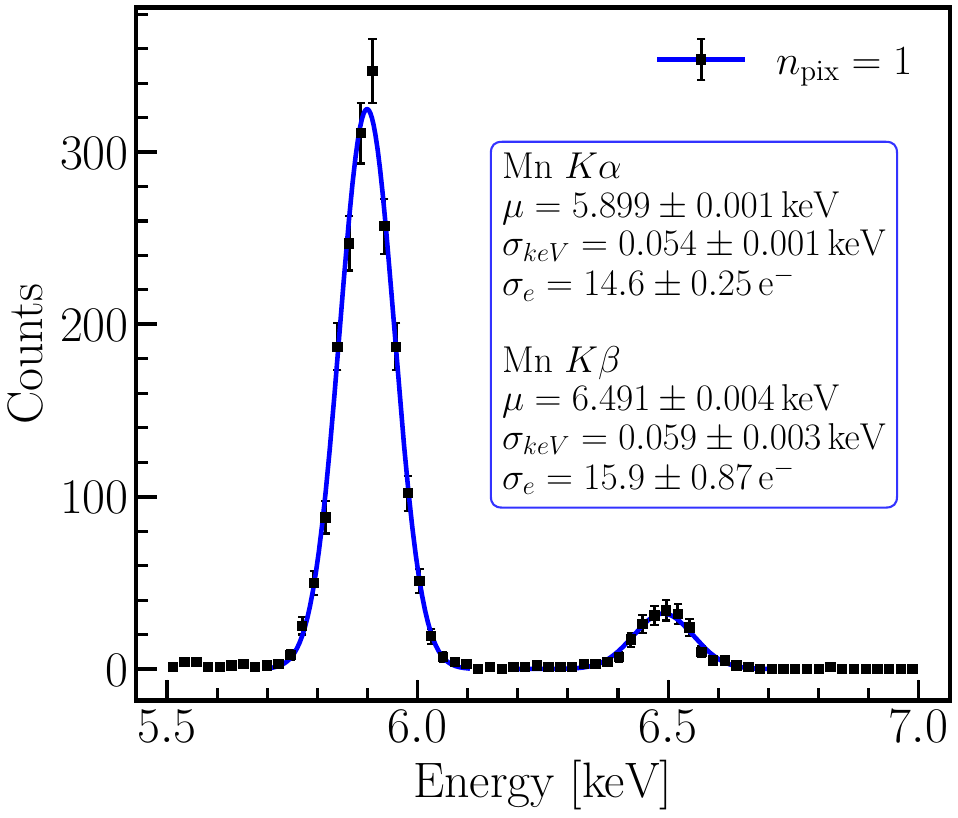}
    \caption{
        Reconstructed $^{55}$Fe spectrum for single-pixel events ($n_{\mathrm{pix}} = 1$). 
        Black points show the data, and the blue curve shows the best-fit model to the Mn $K\alpha$ and $K\beta$ lines. 
        The fit parameters, including the peak centroids and energy resolution, are shown in the inset.
        }
     \label{fig:100-sample-peaks}
\end{figure}

For the front-illuminated device used in this work, $^{55}$Fe interactions occur within the first $\sim$30~$\mu\mathrm{m}$ of the front surface.
At these interaction depths, charge diffusion is minimal, and charge is typically collected in a single pixel or a small number of neighboring pixels.
To evaluate the energy response of the SiSeRO CCD, we restrict our analysis to single-pixel events, minimizing the effect of charge sharing among neighboring pixels on the reconstructed peak width.

After averaging the NDR samples and applying a row-wise overscan subtraction, the images are processed to extract X-ray events.
Events are reconstructed from the corrected image, and only single-pixel events are retained to construct the spectrum.
The energy scale is calibrated by anchoring the photoabsorption peaks to the Mn~$K_{\alpha}$ and Mn~$K_{\beta}$ emission lines.
The resulting spectrum is shown in Fig.~\ref{fig:100-sample-peaks}.
Each peak is fit with a single Gaussian distribution, where the width of the distribution, $\sigma$, corresponds to the energy resolution. 
To convert to units of electrons, we use an average ionization energy of $\sim 3.74$~eV/$e^{-}$, which is consistent for silicon cooled to 130~K~\cite{rodrigues2021absolute, ramanathan_ionization, damic-m_collaboration_precision_2022}.

For single-pixel X-ray events, we measure an energy resolution of  128~eV FWHM (corresponding to $\sigma = 54 \pm 0.9$~eV, or $14.6 \pm 0.25~e^{-}$) for the Mn~$K_{\alpha}$ emission line. 
This resolution demonstrates that the SiSeRO amplifier achieves performance close to the intrinsic charge resolution expected for silicon CCDs at 5.9~keV.

\subsection{Muon-Based Diffusion Calibration}

To empirically calibrate the relationship between transverse charge spread and interaction depth in the sensor, through-going cosmic muons are used as a uniform probe of charge transport throughout the fully depleted volume.
Cosmic muons are minimally ionizing particles that traverse the fully depleted CCD in nearly straight lines, depositing charge continuously along their path~\cite{damic-low-mass-search, CONNIE_2019}. 
This provides a uniform charge source that samples the full sensor depth within a single event. 
Because charge carriers generated deeper in the bulk must drift longer distances to the collection gates, they undergo increased lateral diffusion, resulting in a larger transverse spread. 
This monotonic relationship between diffusion and drift distance allows muon tracks to serve as effective probes of charge transport~\cite{holland2003fully, CONNIE_2019, damic-low-mass-search, background-at-snolab}.

The calibration was constructed from $90$ through-going muon tracks selected after quality cuts to reject fragmented clusters and poorly reconstructed events.
Clean muon clusters are reconstructed as sets of pixels $(x_i, y_i)$ with signal weights $E_i$, and a charge-weighted principal component analysis (PCA) defines local along-track ($s$) and transverse ($t$) coordinates.
Each track is oriented such that the narrower end corresponds to the collection side and the broader end corresponds to the back side of the sensor. 
The track extent is divided into ten equal bins and mapped onto the sensor thickness,
\begin{equation}
z = \frac{s - s_{\min}}{s_{\max} - s_{\min}} \, d,
\end{equation}
with $d = 725~\mu\mathrm{m}$.

\begin{figure}[t]
\centering
\includegraphics[width=1\linewidth]{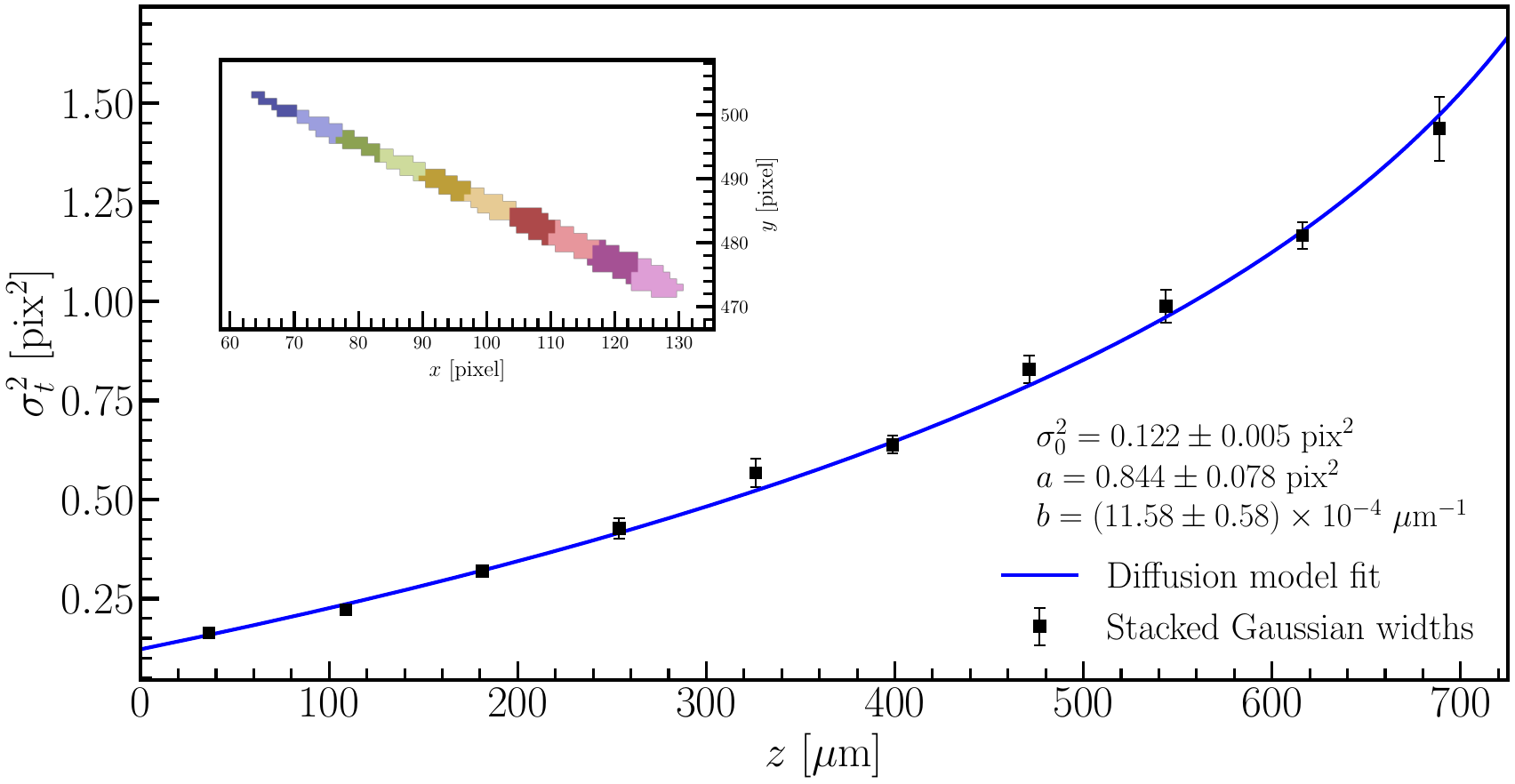}
\caption{
Diffusion variance $\sigma_t^2$ as a function of nominal depth $z$ extracted from through-going cosmic muons.
Points show the Gaussian widths (squared) obtained from stacked transverse charge distributions in each depth bin, with error bars corresponding to the statistical uncertainties of the fits.
The solid curve shows the best-fit diffusion model (see Eq.~\ref{eqn: diffusion}).
Inset: example through-going muon track used in the analysis, with colors indicating the ten depth bins used to construct the diffusion calibration.
The observed increase of $\sigma_t^2$ with depth reflects the growth of lateral charge diffusion with drift distance in the fully depleted sensor.
}
\label{fig:muon_diffusion_calibration}
\end{figure}

For each nominal depth bin, the transverse coordinates of all pixels in that bin are combined into a charge-weighted histogram, where each pixel contributes at its transverse position with weight equal to its signal. 
A Gaussian is fit to each stacked histogram, and the fitted Gaussian width is taken as the effective transverse spread for that depth interval. 
The extracted widths are converted to variances and fit with
\begin{equation}
\sigma_t^2(z) = \sigma_0^2 - a \ln(1 - b z).
\label{eqn: diffusion}
\end{equation}
The resulting calibration curve is shown in Fig.~\ref{fig:muon_diffusion_calibration}, where the measured transverse variances exhibit the expected monotonic increase with nominal depth.
This calibration is therefore interpreted as an empirical detector response relating nominal depth to transverse spread.
The observed depth dependence is consistent with full depletion of the sensor and indicates that charge generated throughout the bulk is efficiently transported to the collection region.

\subsection{Am-241 Measurement}

To illustrate the response of the thick, fully depleted SiSeRO-CCD to higher-energy X-rays, we exposed the sensor to an $^{241}$Am source.
Unlike the $^{55}$Fe measurement, which probes charge generated near the front surface of the CCD, the $^{241}$Am source emits higher-energy photons that interact over a broad range of depths in the 725~$\mu$m fully depleted silicon substrate. 
The reconstructed Np fluorescence complexes between 9--26~keV, together with the visible 59.5~keV $\gamma$ emission, probe detector response across the full sensor volume.

The $^{241}$Am source emits a prominent 59.5~keV $\gamma$-ray, along with a series of Np L-shell fluorescence X-rays produced following the decay of the daughter $^{237}$Np atom~\cite{Lepy_2008}. 
The $\alpha$ emission is attenuated before reaching the CCD by a thin kapton layer.
Additional spectral features arise from the source capsule construction, which includes gold and silver foils~\cite{EckertZieglerCatalog2007}. 
These higher-energy emissions have longer absorption lengths in silicon and produce charge distributions spanning multiple pixels due to diffusion. 
Their reconstruction is consistent with efficient transport and collection of charge generated throughout the thick silicon substrate.

Data for this measurement were acquired using single-sample readout, with short sequential exposures to limit event pile-up.
Before data processing, a row-wise overscan subtraction was applied to each image to remove the electronic baseline.
Reconstructed charge clusters were calibrated to energy by anchoring digital units to the X-ray peaks between 9--26~keV.
Given the increased interaction depth of the source emissions, charge generated by these higher energy X-rays is broadly diffused across multiple pixels.
To generate a clean $^{241}$Am spectrum, cluster selection was limited to those that exhibited a circular profile, reducing contamination from overlapping events and extended charge clusters.

\begin{figure}[t]
  \centering
  \includegraphics[width=\linewidth]{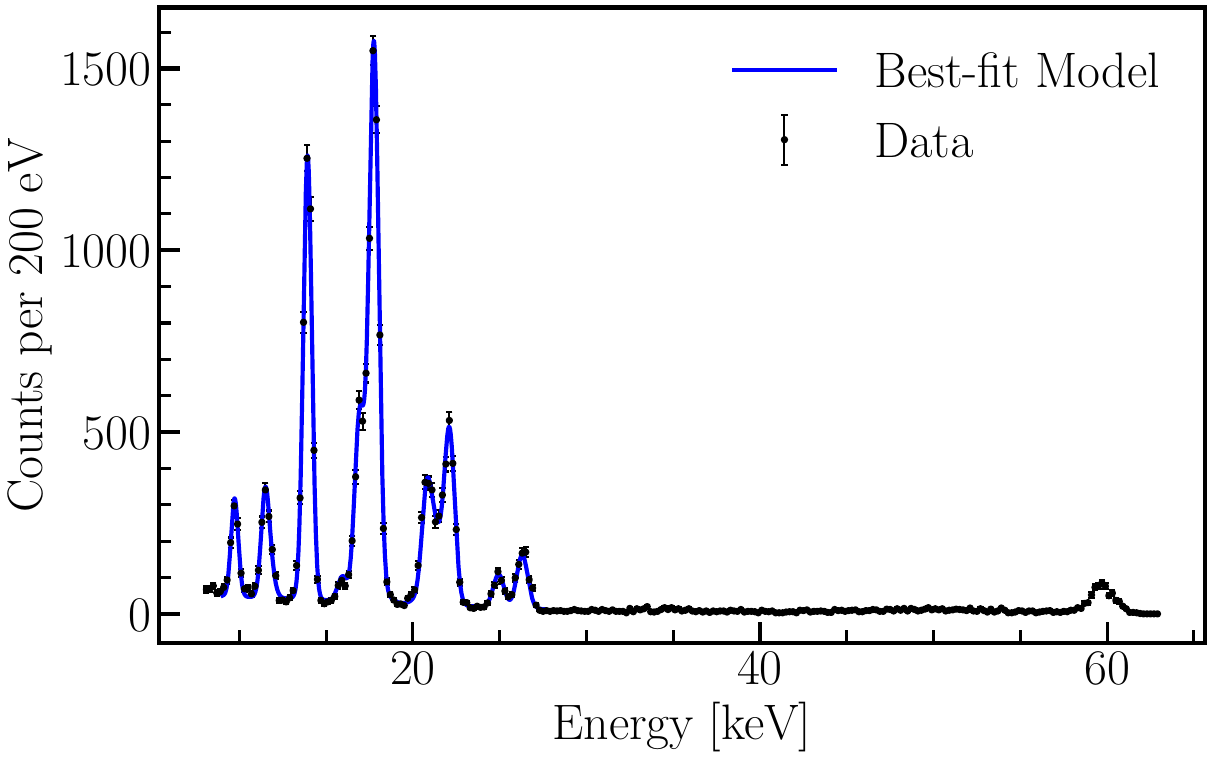}
  \caption{
    Reconstructed $^{241}$Am energy spectrum measured with the SiSeRO detector. 
    Black points show the data (counts per 200~eV bin), and the blue curve shows the best-fit model. 
    The model describes the fluorescence lines in the 9--26~keV range with an energy-dependent resolution and linear background, and is used to establish the energy calibration. 
    The 59.5~keV $\gamma$ line is shown but excluded from the fit.
    }
  \label{fig:am241}
\end{figure}

The resulting spectrum and fit, shown in Fig.~\ref{fig:am241}, is modeled using an empirical line model that includes the dominant emission features expected from $^{241}$Am and characteristic fluorescence lines from the source capsule: 
(i) the 26.3~keV $^{241}$Am emission line, 
(ii) the Np L-shell fluorescence lines (13--21~keV), and 
(iii) Au L fluorescence and Ag K fluorescence from the source capsule.
The 59.5~keV $\gamma$-ray is visible in the spectrum but is not included in the spectral fit due to event pile-up and broader charge diffusion at this energy.

The individual line energies were fixed from tabulated atomic data, while the amplitudes of the Ag and Au fluorescence lines are free parameters in the fit.
An energy-dependent Gaussian resolution,
\begin{equation}
\sigma_{res}(E)=\sigma_{res,0} + k_E E,
\end{equation}
accounts for the degradation in resolution with increasing energy.
Here $\sigma_{\mathrm{res},0}$ is the baseline resolution term, $k_E$ describes the energy dependence of the peak width, and $E$ is the X-ray energy.
A linear background term absorbs the smooth continuum between peaks.

The reconstruction of both the fluorescence lines and the 59.5~keV $\gamma$ emission is consistent with charge collection throughout the thick silicon substrate.
While the $^{55}$Fe measurements probe near-surface interactions, the successful detection of higher-energy Am-241 emissions confirms efficient charge collection throughout the full substrate depth. 
These measurements are consistent with the performance expected for a fully depleted silicon CCD~\cite{holland2003fully}, including efficient charge collection across the thick substrate.
These results support the viability of SiSeRO for X-ray spectroscopy applications, in addition to its demonstrated capability for single-electron-resolved photon counting in faint-signal astronomical imaging.

\section{Discussion and Conclusion}

We have evaluated the X-ray response of the fully depleted, p-channel SiSeRO-CCD using $^{55}$Fe and $^{241}$Am radiation sources. 
The $^{55}$Fe measurement provides a direct characterization of the detector energy resolution at 5.9~keV, yielding $54 \pm 0.9$~eV ($14.6 \pm 0.24~e^{-}$) for single-pixel events and demonstrating that the SiSeRO amplifier preserves the intrinsic charge resolution of the CCD while operating in a multi-sample non-destructive readout mode. 
Measurements with the $^{241}$Am source extend this validation to higher-energy photons that interact throughout the 725~$\mu$m fully depleted substrate, while the muon-derived diffusion calibration is consistent with efficient charge transport throughout the detector bulk.

Taken together, these results show that the SiSeRO-CCD combines sub-electron charge sensitivity with a thick, fully depleted silicon detection volume, enabling low-noise charge measurement with efficient bulk transport and collection. 
These properties support the broad-range X-ray response demonstrated here and are well matched to the requirements of optical and near-infrared observations in future ground- and space-based instruments. 
In this context, SiSeRO provides a path toward astronomical imagers that achieve photon-counting sensitivity while operating at higher readout speeds than conventional floating-gate non-destructive architectures.

\section*{Disclosures}

The authors declare no conflicts of interest. During the preparation of this work, the authors made limited use of generative AI tools for code prototyping, figure preparation, and language refinement. The authors reviewed and validated all scientific content.

\section*{Code and Data Availability}
The data and analysis code used in this study may be obtained from the corresponding author upon reasonable request and subject to institutional approval.

\acknowledgments
Fermilab is managed by Fermi Forward Discovery Group, LLC, acting under Contract No.\ 89243024\allowbreak{}CSC000002 for the U.S. Department of Energy. This work was partially supported by NASA SAT award No.~80NSSC25K7508.

The authors thank Brenda A. Cervantes-Vergara and Javier Tiffenberg for useful discussions regarding charge diffusion and depth reconstruction in silicon CCDs.

\bibliographystyle{spiebib}
\bibliography{report}

\end{spacing}
\end{document}